%% file: 00_main.tex
\documentclass[conference]{IEEEtran}
\IEEEoverridecommandlockouts

\usepackage{cite}
\usepackage{hyperref}
\usepackage{amsmath,amssymb,amsfonts}
\usepackage{algorithmic}
\usepackage[pdftex]{graphicx}
\usepackage{subfigure}
\usepackage{textcomp}
\usepackage{booktabs}
\usepackage{flushend}
\usepackage{enumitem}
\usepackage{multirow}
\usepackage{siunitx}
\usepackage{url}
\usepackage[acronym,hyperfirst=false]{glossaries}
\def\BibTeX{{\rm B\kern-.05em{\sc i\kern-.025em b}\kern-.08em
    T\kern-.1667em\lower.7ex\hbox{E}\kern-.125emX}}

\newacronym{cnn}{CNN}{Convolutional Neural Network}
\newacronym{gan}{GAN}{Generative Adversarial Network}
\newacronym{tp}{TP}{True Positives}
\newacronym{fp}{FP}{False Positives}
\newacronym{tn}{TN}{True Negatives}
\newacronym{fn}{FN}{False Negatives}
\newacronym{tpr}{TPR}{True Positive Rate}
\newacronym{fpr}{FPR}{False Positive Rate}
\newacronym{dtw}{DTW}{Dynamic Time Warping}
\newacronym{vad}{VAD}{Voice Activity Detector}
\newacronym{tts}{TTS}{Text-to-speech}
\newacronym{vc}{VC}{Voice Conversion}
\newacronym{asv}{ASV}{Automatic Speaker Verification}
\newacronym{mos}{MOS}{Mean Opinion Score}
\newacronym{roc}{ROC}{Receiver Operating Characteristic}
\newacronym{auc}{AUC}{Area Under the Curve}

\def\BibTeX{{\rm B\kern-.05em{\sc i\kern-.025em b}\kern-.08em
    T\kern-.1667em\lower.7ex\hbox{E}\kern-.125emX}}
\begin{document}

\title{TIMIT-TTS: a Text-to-Speech Dataset for Multimodal Synthetic Media Detection}
\author{\IEEEauthorblockN{
Davide Salvi\IEEEauthorrefmark{1},
Brian Hosler\IEEEauthorrefmark{2},
Paolo Bestagini\IEEEauthorrefmark{1},
Matthew C. Stamm\IEEEauthorrefmark{2},
Stefano Tubaro\IEEEauthorrefmark{1}
}
\IEEEauthorblockA{\IEEEauthorrefmark{1}\textit{Dipartimento di Elettronica, Informazione e Bioingegneria (DEIB), Politecnico di Milano}}

\IEEEauthorblockA{\IEEEauthorrefmark{2}\textit{Department of Electrical and Computer Engineering, Drexel University, Philadelphia, PA, USA}}
}

\maketitle

\begin{abstract}
With the rapid development of deep learning techniques, the generation and counterfeiting of multimedia material are becoming increasingly straightforward to perform.
At the same time, sharing fake content on the web has become so simple that malicious users can create unpleasant situations with minimal effort.
Also, forged media are getting more and more complex, with manipulated videos (e.g., deepfakes where both the visual and audio contents can be counterfeited) that are taking the scene over still images.
The multimedia forensic community has addressed the possible threats that this situation could imply by developing detectors that verify the authenticity of multimedia objects.
However, the vast majority of these tools only analyze one modality at a time.
This was not a problem as long as still images were considered the most widely edited media, but now, since manipulated videos are becoming customary, performing monomodal analyses could be reductive.
Nonetheless, there is a lack in the literature regarding multimodal detectors (systems that consider both audio and video components). This is due to the difficulty of developing them but also to the scarsity of datasets containing forged multimodal data to train and test the designed algorithms.

In this paper we focus on the generation of an audio-visual deepfake dataset.
First, we present a general pipeline for synthesizing speech deepfake content from a given real or fake video, facilitating the creation of counterfeit multimodal material.
The proposed method uses Text-to-Speech (TTS) and Dynamic Time Warping (DTW) techniques to achieve realistic speech tracks.
Then, we use the pipeline to generate and release TIMIT-TTS, a synthetic speech dataset containing the most cutting-edge methods in the TTS field.
This can be used as a standalone audio dataset, or combined with DeepfakeTIMIT and VidTIMIT video datasets to perform multimodal research.
Finally, we present numerous experiments to benchmark the proposed dataset in both monomodal (i.e., audio) and multimodal (i.e., audio and video) conditions.
This highlights the need for multimodal forensic detectors and more multimodal deepfake data.
\end{abstract}


\input{01_introduction}
\input{02_background}
\input{03_dataset_generation}
\input{04_dataset_description}
\input{05_results}

\input{06_conclusion}

\section*{Acknowledgment}

Research was sponsored by the Army Research Office and was accomplished under Cooperative Agreement Number W911NF-20-2-0111, the Defense Advanced Research Projects Agency (DARPA) and the Air Force Research Laboratory (AFRL) under agreement numbers FA8750-20-2-1004 and HR001120C0126, and by the National Science Foundation under Grant No. 1553610. The U.S. Government is authorized to reproduce and distribute reprints for Governmental purposes notwithstanding any copyright notation thereon. The views and conclusions contained herein are those of the authors and should not be interpreted as necessarily representing the official policies or endorsements, either expressed or implied, of DARPA and AFRL, the Army Research Office, the National Science Foundation, or the U.S. Government. This work was partly supported by the PREMIER project, funded by the Italian Ministry of Education, University, and Research within the PRIN 2017 program.

{\small
\bibliographystyle{unsrt}
\bibliography{bibliography}
}

\end{document}

%% file: 01_introduction.tex
\section{Introduction}
\label{sec:introduction}

In recent years, deep learning technologies have grown fast and relentlessly.
New scenarios that were only imaginable a few years ago are now possible, and others will arise soon.
For example, virtual assistants, natural language processing and visual recognition algorithms have become commonplace and help simplify several daily tasks.
New categories of media are also born, like deepfakes: videos obtained through AI-driven technologies capable of synthesizing a target person's identity or biometric aspects.
Deepfake generation technologies can create exciting unexplored scenarios, but can also lead to dangers and threats when misused.
For example, deepfake techniques allow to generate video content representing a victim in deceiving situations and/or behaviors, which can lead to frauds, scam cases and fake news spreading~\cite{deepfakenyt, Wired, cnn_deepfake}.
This menace cannot be ignored, as we have reached a point where we are no longer able to always distinguish real media from artificially generated ones~\cite{fakenewscientist, NPR}.

The scientific community has embraced this problem and has started working in several directions to moderate deepfake online spread considering both audio and video threats~\cite{verdoliva2020media}.
For instance, international challenges have been organized to make people aware of the importance of fighting deepfake misuse.
To this purpose the DFDC challenge~\cite{dolhansky2020deepfake} focused on video deepfake detection, while ASVspoof~\cite{todisco2019asvspoof, yamagishi2021asvspoof} and ADD~\cite{yi2022add} challenges have been proposed in the audio field.
Furthermore, part of the research community has focused on releasing deepfake datasets to help develop forensic detectors.
This is the case of Faceforensics++~\cite{rossler2019faceforensics++} and DeepfakeTIMIT~\cite{korshunov2018deepfakes} for videos, as well as WaveFake~\cite{frank2021wavefake} for audio.
Most important, multimedia forensics researchers have developed several detectors to discriminate deepfakes from pristine media.
These leverage different characteristics, ranging from low-level artifacts left by the generators~\cite{durall2019unmasking, guarnera2020deepfake} to more semantic aspects~\cite{li2018ictu, agarwal2020detecting}.

Despite the considerable effort put into fighting deepfakes, a common trait of the developed detectors is that they primarily focus on monomodal analysis: they consider either the audio or video deepfake detection problem separately.
Since videos usually come with audio tracks, and both the visual and audio content are subject to editing, performing a joint audio-visual multimodal analysis should be the preferred option.
However, only a few approaches have been proposed to perform multimodal detection leveraging inconsistencies or traces orthogonal to different modalities to identify counterfeit materials.
For example,~\cite{hosler2021deepfakes} exploits the inconsistencies between emotions conveyed by audio and visual modalities to perform a joint audio-visual deepfake detection.
The authors of~\cite{lomnitz2020multimodal} incorporate temporal information from series of images, audio and video data to provide a multimodal deepfake detection approach.
Alternatively, the authors of~\cite{khalid2021evaluation} show that combining audio and video baselines in an ensemble-based method provides better detection performance than a monomodal system.

The main reason for the lack of multimodal forensic systems for deepfake detection is the scarcity of data to train and test them.
Most of these systems are data-driven and require a large amount of data to be trained. Still, most of the deepfake datasets proposed in the literature are monomodal. There is a dearth of challenging fake video datasets that also contain fake audio, making it tough to develop multimodal systems.

In this paper we address the problem of lack of deepfake multimodal data by focusing on three main contributions
\begin{itemize}
    \item We propose a general pipeline to turn a monomodal video deepfake dataset from the literature into a multimodal audio-visual deepfake dataset.
    \item We apply the proposed pipeline to the VidTIMIT~\cite{sanderson2002vidtimit} and DeepfakeTIMIT~\cite{korshunov2018deepfakes} datasets in order to build and release the novel multimodal TIMIT-TTS deepfake dataset containing almost \num{80 000} tracks.
    \item We benchmark the generated dataset by running a series of deepfake detection baselines that highlight the main challenges for future research.
\end{itemize}

The rationale behind the proposed pipeline is that realistic deepfake video datasets have been proposed in the literature, but these do not contain accompanying deepfake audio.
Therefore, we present a technique to generate a synthetic speech track for a given input video.
This approach allows us to generate fake audio content starting from any video containing speech, considering the most advanced state-of-the-art \gls{tts} systems.
Once generated, the synthetic track can be paired with the input video and, depending on the authenticity of the latter, an audio-only or an audio-visual deepfake is generated.
Our pipeline thus provides a viable solution for making counterfeit multimodal materials, which is in general complex to perform.

To showcase the actual feasibility of the proposed deepfake generation approach, we apply it to the VidTIMIT dataset~\cite{sanderson2002vidtimit} and DeepfakeTIMIT dataset~\cite{korshunov2018deepfakes}.
The former contains audio-video recordings of \num{43} people speaking.
The latter is a video deepfake version of the former.
By generating synthetic speech for both video datasets, we end up with the proposed TIMIT-TTS, a synthetic speech dataset built using state-of-the-art \gls{tts} techniques.
On the one hand, TIMIT-TTS can be used as a standalone audio dataset to test the developed speech deepfake detectors, as it contains the most cutting-edge methods in the synthetic speech synthesis field.
On the other hand, TIMIT-TTS can also be combined with VidTIMIT and DeepfakeTIMIT to provide multimodal audio-video deepfake data, which is an overlooked aspect in the current literature.

Finally, we run a series of tests to provide some information on the challenges proposed by this new multimodal dataset.
We adopt the video deepfake detector proposed in~\cite{bonettini2021video} and the audio deepfake detector proposed in~\cite{tak2021end} to analyze videos and audio tracks in both monomodal and multimodal fashion.
Results confirm that multimodal deepfake analysis should be preferred and show that audio deepfake attribution is an interesting topic for further research.

The rest of paper is structured as follows.
Section~\ref{sec:background} recap the motivations behind our work, and provides the reader with some useful background on generation and detection methods for speech deepfakes.
Section~\ref{sec:pipeline} describes the proposed generation pipeline for the deepfake audio tracks, and provides an overview of the considered \gls{tts} synthesis algorithms.
Section~\ref{sec:description} explains the structure of the released TIMIT-TTS dataset.
Section~\ref{sec:results} presents the results of the analysis conducted on the released data.
Finally, Section~\ref{sec:conclusion} concludes the paper along with a brief discussion of possible future work.

%% file: 02_background.tex
\section{Overview and Background}
\label{sec:background}

This section provides the reader with some helpful background information needed to understand the primary rationale behind our proposal.
First, we highlight the need for a multimodal deepfake dataset with a particular focus on synthetic speech generation, as the one proposed in this paper.
Then, we provide a quick overview of synthetic speech generation and detection techniques, which are at the base of our proposed dataset and benchmarking work.

\subsection{Motivations}

Numerous deepfake datasets have been proposed in recent years, both in the audio and video case, significantly pushing research towards developing new methods for recognizing counterfeit material.
The publication of these sets leads to designing more innovative and effective detectors since they provide new data on which to train and test them.
However, most of the presented datasets focus only on one modality at a time, resulting in valuable data for producing monomodal detectors but not relevant for multimodal methods.
Indeed, to train and test multimodal detectors, there is a need for data that are altered in all the considered aspects (e.g., both video and audio).
The lack of this data is one of the main reasons behind the lack of multimodal detectors investigations and is the primary motivation behind this work.

Recently, two multimodal deepfake datasets have been proposed, both containing counterfeit audio and video. These are DFDC~\cite{dolhansky2020deepfake} and FakeAVCeleb~\cite{khalid2021fakeavceleb}.
Although these propose a solution to the abovementioned problem, we cannot define either of these as complete, especially from an audio point of view.
On one side, DFDC does not provide labels as to which of the audio or visual components are fake, but the content is labeled as fake when at least one of the two modalities is counterfeit. Therefore we do not have sufficient information to perform tests on different scenarios (e.g., fake audio and real video or vice versa) and investigate which aspects the detector leverages to discriminate between real and altered data.
On the other hand, the multimodal deepfakes contained in FakeAVCeleb are generated overlooking the audio modality. All the fake audio tracks are synthesized using the same \gls{tts} algorithm, and none of them is synchronized with the corresponding video. This results in a lack of both variety and realism in the released data.

In this paper we address both these problems by proposing a pipeline to generate multimodal deepfake datasets starting from video deepfake ones.
Indeed, we can make use of the good quality video datasets proposed in the literature and automatize the generation of realistic speech tracks to be synchronized and matched with these videos.
We use this pipeline to release TIMIT-TTS, a synthetic speech dataset that overcomes the abovementioned limitations.
The released data include synthetic speech generated from \num{12} different \gls{tts} systems, providing an overview of the most advanced techniques in state-of-the-art.
The tracks have also been synchronized with the related videos using a \gls{dtw} technique, producing highly realistic content, as shown in Section~\ref{sec:results}.
These two aspects, namely the variety of speech generation algorithms and time warping, fill a gap that is present in state-of-the-art multimodal deepfake datasets.
Given the variety of generation methods it includes, TIMIT-TTS can be used both as a standalone synthetic audio dataset and to perform multimodal deepfake studies, used in conjunction with other well-established deepfake video datasets.

\subsection{Speech Deepfake Generation methods}

Deepfake content generation techniques are becoming increasingly simple to use and the data they produce are getting more and more realistic. In some cases, the generated synthetic material is so lifelike that it is difficult to discern from an authentic one~\cite{fakenewscientist}.
Although this is true for both audio and video data, here we focus on the generation methods of speech deepfakes, which are the main subject of study in this paper.

As far as synthetic speech data generation is concerned, techniques can be broadly split into two main families: \gls{tts} methods and \gls{vc} methods.
The difference between these two kinds of techniques is mainly the input of the generation system.
\gls{tts} algorithms produce speech signals starting from a given text.
Conversely, \gls{vc} methods take a speech signal as input and alter it by changing its style, intonation or prosody, trying to mimic a target voice.

Regarding \gls{tts} methods, a long history of classical techniques based on vocoders and waveform concatenation has been proposed in the literature~\cite{klatt1987review}.
However, the first modern breakthrough that significantly outperformed all the classical methods was introduced by WaveNet~\cite{van2016wavenet}, a neural network for generating raw audio waveforms capable of emulating the characteristics of many different speakers. 
This network has been overtaken over the years by other systems~\cite{kuchaiev2019nemo, wang2017tacotron}, which made the synthesis of highly realistic artificial voices within everyone's reach.

Most \gls{tts} systems follow a two-step approach. First, a model generates a spectrogram starting from a given text. Then, a vocoder synthesizes the final audio from the spectrogram. This approach allows combining different vocoders for the same spectrogram generator and vice versa.
Alternatively, some end-to-end models have been proposed, which generate speech directly from the input text~\cite{Hayashi2020Espnet}.

Considering \gls{vc} algorithms, the earliest models were based on spectrum mapping using parallel training data~\cite{toda2007voice, desai2010spectral}.
However, most of the current approaches are \gls{gan}-based~\cite{kameoka2018stargan, kaneko2019cyclegan}, allowing to learn a mapping from source to target speaker without relying on parallel data.

In this work we only consider \gls{tts} methods as they are more investigated in the literature and allow us to build a more various dataset.
Indeed, as we aim for a fully automated and high-quality pipeline, \gls{tts} methods are less prone to errors compared to \gls{vc} techniques.
Moreover, in principle \gls{tts} methods allow for easy editing of even just one single word within a speech.
Furthermore, \gls{tts} realistic voice styles are typically easier to tune compared to \gls{vc} techniques whose fine-tuning if often challenging and time-consuming.
Nevertheless, also \gls{vc} methods are worth further studies and will be the subject of future versions of this dataset.


\subsection{Speech Deepfake detection methods}

The speech deepfake detection task consists in determining whether a given speech track $x$ is authentic from a real speaker or has been synthetically generated.
Recently, this has become a hot topic in the forensic research community, trying to keep up with the rapid evolution of counterfeiting techniques~\cite{lyu2020deepfake}.

In general, speech deepfake detection methods can be divided into two main groups based on the aspect they leverage to perform the detection task.
The first focuses on low-level aspects, looking for artifacts introduced by the generators at the signal level. In contrast, the second focuses on higher-level features representing more complex aspects as the semantic ones.

As an example of artifacts-based approaches, \cite{wang2011channel} aims to secure \gls{asv} systems against physical attacks through channel pattern noise analysis.
In~\cite{malik2019securing}, the authors assume that a real recording has more significant non-linearity than a counterfeit one, and they use specific features, such as bicoherence, to discriminate between them.
Bicoherence is also employed in~\cite{borrelli2021synthetic} along with several features based on modeling speech as an auto-regressive process. The authors investigate whether these features complement and benefit each other.
Alternatively, the authors of~\cite{tak2021end} propose an end-to-end network to spot synthetic speech.

On the other hand, detection approaches that rely on semantic features are based on the hypothesis that deepfake generators can synthesize low-level aspects of the signals but fail in reproducing more complex high-level features.
For example, \cite{sahidullah2015comparison} exploits the deepfake detection task by relying on classic audio features inherited from the music information retrieval community.
The authors of~\cite{conti2022deepfake} exploit the lack of emotional content in synthetic voices generated via \gls{tts} techniques to recognize them.
Finally, in~\cite{attorresi2022prosody} \gls{asv} and prosody features are combined to perform synthetic speech detection.

%% file: 03_dataset_generation.tex
\section{Dataset Creation Methodology}
\label{sec:pipeline}

This section presents the methodology we propose to generate a deepfake speech track for a given input video, being the video real or fake. In doing so, we also detail all the implemented \gls{tts} systems used to synthesize the signals and the techniques applied to post-process them. This is the pipeline we follow to generate the proposed dataset.

\subsection{Generation pipeline}

\begin{figure}
    \centering
    \includegraphics[width=.9\columnwidth]{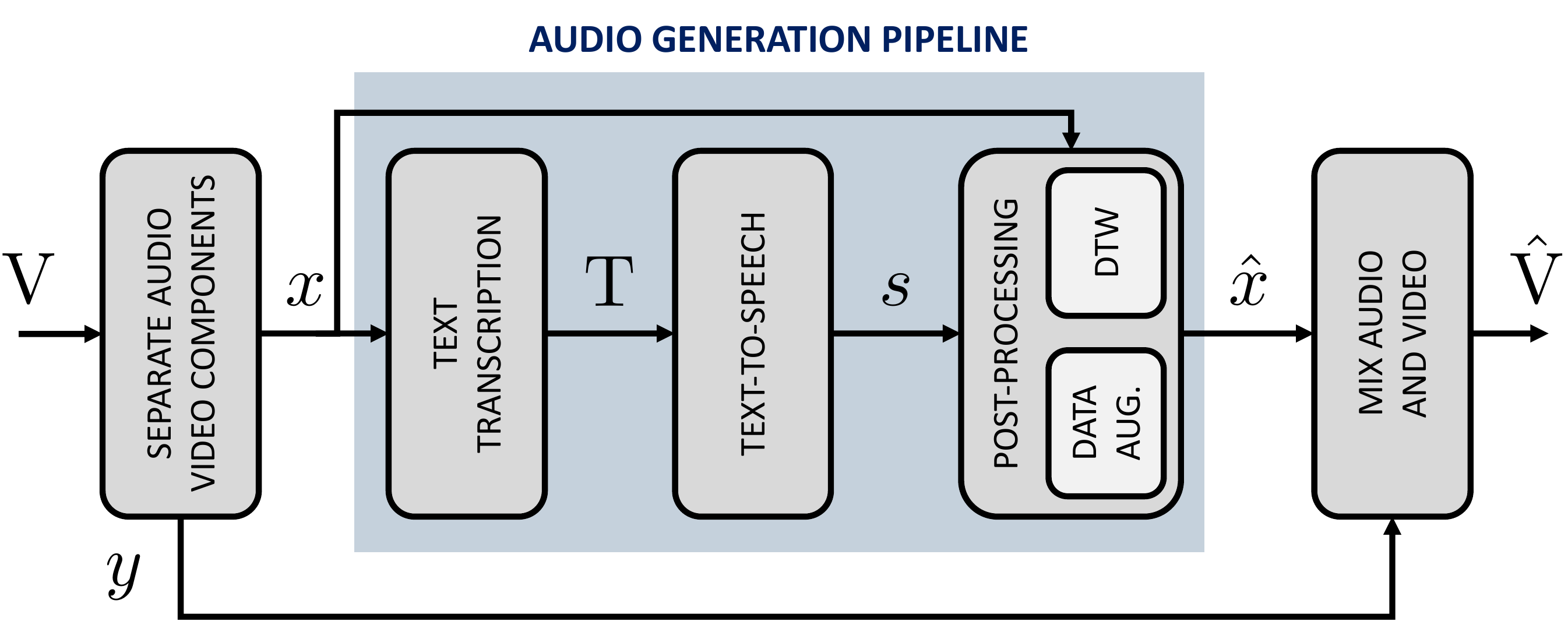}
    \caption{Pipeline of the proposed generation method.}
    \label{fig:pipeline}
\end{figure}

The proposed pipeline to generate a synthetic speech track for a given video comprises several steps, as it is shown in Figure~\ref{fig:pipeline}.
The input to the whole process consists of a video $\mathrm{V}$ that represents a speaking person.
Here we consider a video $\mathrm{V}$ as a multimedia object composed of both an audio speech content $x$ and a visual component $y$ depicting a person's face, as in 
\begin{equation}
    \mathrm{V} = x \oplus y ,
\end{equation}
where $\oplus$ is the mixing operation between the audio and visual signals.
Our final goal is to produce a forged video $\hat{\mathrm{V}}$ containing the same visual subject as $\mathrm{V}$ but where the speech track $\hat{x}$ is a deepfake synthetically generated.
To summarize, we can write
\begin{align}
    \hat{\mathrm{V}} &= \Lambda(\mathrm{V}), \\
    \hat{x} \oplus y &= \Lambda(x \oplus y) ,
\end{align}
where $\Lambda(\cdot)$ indicates the complete pipeline we propose.

To achieve our goal, the first operation we perform is to split $\mathrm{V}$ into its components $x$ and $y$.
The speech track $x$ becomes the input of the \textit{audio generation pipeline}, which outputs its synthetic counterpart $\hat{x}$.
This segment is composed of three main blocks.
The first is a speech-to-text algorithm, which transcribes the speech content of $x$ into the text $\mathrm{T}$.
The second block is a \gls{tts} algorithm that produces a synthetic audio track $s$ from a given string $\mathrm{T}$, as in
\begin{equation}
    s = \Gamma(\mathrm{T}) ,
\end{equation}
where $\Gamma(\cdot)$ is one of the \gls{tts} systems presented later in this section.
Finally, the third block consists of a post-processing step, which takes the generated track $s$ as input and outputs its processed version $\hat{x}$, which is more realistic and challenging to discriminate for deepfake detectors.
$\hat{x}$ is the deepfake version of the $x$ input speech track.

Two different post-processing techniques are implemented in our pipeline, which can be applied individually or together. In case neither is applied, we output the clean signal $\hat{x} = s$.
The first technique is speech-to-speech synchronization based on \gls{dtw}.
Since the goal of the proposed system is to generate a fake speech track for a given video, we need the synthesized audio to be synchronized with the video itself.
Without performing the alignment, the synthetic track $\hat{x}$ will have a different temporal trend from the input audio $x$ and the corresponding video $\mathrm{V}$. This results in a deepfake that is very easy to detect for all the systems trained to analyze the discrepancies in time between the audio and video modalities.
This pipeline step takes as input the two audio signals $s$ and $x$ and performs time warping on the former by mapping it to the latter. We do so through the alignment algorithm presented later in this section.
The output track $\hat{x}$, being synchronized with $x$, is also synchronized with the input video $\mathrm{V}$. 

The second block of the post-processing step consists of data augmentation.
Here we apply several algorithms, including noise injection, pitch shifting and lossy compression, to make the generated data more challenging to discriminate for those deepfake detectors that are not robust to such operations.
In fact, this kind of processing operations hinder the traces that \gls{tts} algorithms could leave, making the generated data tougher to identify.
Finally, once the audio track $\hat{x}$ has been obtained, we mix it with the input video $y$ generating a new multimodal deepfake content $\hat{\mathrm{V}} = \hat{x} \oplus y$.
Depending on the authenticity or not of the input video, $\hat{\mathrm{V}}$ will be a mono or multimodal deepfake.

\subsection{Speech synthesis}

In the proposed pipeline, the \gls{tts} block can support multiple speech generation algorithms $\Lambda$.
We did so to add the possibility of generating data with different characteristics, not related to a single algorithm and more representative of the state-of-the-art.
Most of the considered \gls{tts} algorithms follow a two-stage pipeline, while only a few methods have an end-to-end approach, generating speech signals directly from an input text.
In the two-stage case, the first block takes a text as input and generates a spectrogram, while the second is a vocoder that sonifies the output of the first step.
The two blocks are independent from each other and we can potentially use different vocoders for the same spectrogram generator.
Here we consider a \gls{tts} method $\Lambda$ as a fixed pair of generator and vocoder.
Even though this interchangeability allows us to potentially have a large number of methods $\Lambda$, in this study we want to limit the number of vocoders considered.
We do so since we want to keep the differences between the generated speech tracks primarily attributable to the spectrogram generators.
Nevertheless, we believe that the artifacts introduced by the vocoders are a noteworthy aspect and these will be the subject of subsequent versions of this dataset.

Here is a list of the considered spectrogram generators.
\begin{itemize}[leftmargin=*]
\item \textbf{Tacotron}~\cite{wang2017tacotron} is a seq2seq model, which includes an encoder, an attention-based decoder, and a post-processing net. Both the encoder and decoder are based on Bidirectional GRU-RNN. We consider the version implemented in~\cite{tacotron_repo}.
\item \textbf{Tacotron2}~\cite{shen2018natural} has the same architecture as Tacotron but improves its performance by adding a Location Sensitive Attention module to connect the encoder to the decoder.
\item \textbf{GlowTTS}~\cite{kim2020glow} is a flow-based generative model. It searches for the most probable monotonic alignment between text and the latent representation of speech on its own, enabling robust and fast \gls{tts} synthesis.
\item \textbf{FastSpeech2}~\cite{ren2020fastspeech} is composed of a Transformer-based encoder and decoder, together with a variance adaptor that predicts variance information of the output spectrogram, including the duration of each token in the final spectrogram and the pitch and energy per frame.
\item \textbf{FastPitch}~\cite{lancucki2021fastpitch} is based on FastSpeech, conditioned on fundamental frequency contours. It predicts pitch contours during inference to make the generated speech more expressive.
\item \textbf{TalkNet}~\cite{beliaev2020talknet} is consists of two feed-forward convolutional networks. The first predicts grapheme durations by expanding an input text, while the second generates a Mel-spectrogram from the expanded text. 
\item \textbf{MixerTTS}~\cite{tatanov2021mixer} is based on the MLP-Mixer architecture adapted for speech synthesis. The model contains pitch and duration predictors, with the latter being trained with an unsupervised \gls{tts} alignment framework. 
\item \textbf{MixerTTS-X}~\cite{tatanov2021mixer} has the same architecture as MixerTTS but additionally uses token embeddings from a pre-trained language model.
\item \textbf{VITS}~\cite{kim2021conditional} is a parallel end-to-end \gls{tts} method that adopts variational inference augmented with normalizing flows and an adversarial training process to improve the expressive power of the generated speech.
\item \textbf{SpeedySpeech}~\cite{vainer2020speedyspeech} is a student-teacher network capable of fast synthesis, with low computational requirements. It includes convolutional blocks with residual connections in both student and teacher networks and uses a single attention layer in the teacher model.
\item \textbf{gTTS}~\cite{gtts} (\textit{Google Text-to-Speech}) is a Python library and CLI tool to interface with Google Translate’s text-to-speech API. It generates audio starting from an input text through an end-to-end process. 
\item \textbf{Silero}~\cite{Silero_Models} pre-trained enterprise-grade \gls{tts} model that works faster than real-time following an end-to-end pipeline.
\end{itemize}

Here is a list of the considered vocoders.
\begin{itemize}[leftmargin=*]
\item \textbf{MelGAN}~\cite{kumar2019melgan} is a \gls{gan} model that generates audio from mel-spectrograms. It uses transposed convolutions to upscale by the mel-spectrogram to audio. We considered this vocoder to generate speech from Tacotron2, GlowTTS, FastSpeech2, FastPitch, TalkNet, MixerTTS, MixerTTS-X, and SpeedySpeech.
\item \textbf{WaveRNN}~\cite{kalchbrenner2018efficient} is a single-layer recurrent neural network with a dual softmax layer, able to generate audio 4x faster than real-time. We considered this vocoder to generate audio from Tacotron.
\end{itemize}

Most of the models mentioned above follow a deep-learning approach and the data they generate is highly dependent on the one seen during the training phase. 
This also affects the speakers' number and identity that a model supports. In fact, if a system has been trained with numerous speakers, it will also be able to reproduce them at inference time, resulting in a multi-speaker generator. Conversely, if we train a system on one speaker only, it will be able to generate audio only with that tone of voice.

Here is a list of the datasets considered for training the used \gls{tts} methods in order to obtain different voice styles.
\begin{itemize}[leftmargin=*]
\item \textbf{LJSpeech}~\cite{LJspeech} is a dataset containing short audio tracks of speech recorded from a single speaker reciting pieces from non-fiction books.
\item \textbf{LibriSpeech}~\cite{panayotov2015librispeech} is a dataset that contains about \num{1000} hours of authentic speech from more than \num{200} different speakers.
\item \textbf{CSTR VCTK Corpus}~\cite{yamagishi2019cstr} (Centre for Speech Technology Voice Cloning Toolkit) is a dataset that includes speech data uttered by \num{109} native speakers of English with various accents. Each speaker reads about \num{400} sentences from a newspaper and a passage intended to identify the speaker's accent. 
\end{itemize}

Table~\ref{tab:speakers} presents a summary of the datasets used to train each algorithm, together with the implemented number of speakers in TIMIT-TTS. The models trained on LibriSpeech and VCTK support multi-speaker synthesis, while those trained on LJSpeech only support a single speaker, which is an English female voice with an American accent.
For gTTS, no dataset is indicated as it directly interfaces with Google Translate's \gls{tts} API and synthesizes speech using its pre-trained models. This model supports English in 4 different accents (United States, Canada, Australia and India).
It is worth noting that several methods have been trained on LJSpeech, resulting in diverse systems able to generate speech with the same voice.
This allows generating speech data that are not biased by the speaker's identity and that are more difficult to discriminate by deepfake detectors, as shown in Section~\ref{sec:results}.

\begin{table}[]
\centering
\caption{Datasets used to train each \gls{tts} method and considered number of speakers in TIMIT-TTS.}
\label{tab:speakers}
\begin{tabular}{lcc}
\toprule
\textbf{Generator}    & \textbf{Dataset}               & \textbf{Num. Speakers} \\ \midrule
\midrule
gTTS         & //                    & 4             \\
Tacotron     & LibriSpeech           & 8             \\
GlowTTS      & LJSpeech, VCTK        & 9             \\
FastPitch    & LJSpeech, VCTK        & 9             \\
VITS         & LJSpeech, VCTK        & 9             \\
FastSpeech2  & LJSpeech              & 1             \\
MixerTTS     & LJSpeech              & 1             \\
MixerTTS-X   & LJSpeech              & 1             \\
SpeedySpeech & LJSpeech              & 1             \\
Tacotron2    & LJSpeech              & 1             \\
TalkNet      & LJSpeech              & 1             \\
Silero       & LJSpeech              & 1             \\ \bottomrule

\end{tabular}
\end{table}

\subsection{Audio-Video synchronization}

To generate a realistic audio-video deepfake, we need its audio and visual components to be synchronized with each other.
This is crucial as diverse semantic deepfake detectors leverage the inconsistencies between the two modalities to discriminate among authentic and counterfeited media contents~\cite{chugh2020not} and having the two components asynchronous would result in deepfake easy to spot.
To avoid this, we synchronize the generated \gls{tts} track $s$ with the original audio $x$ of the input video.
Since $x$ is aligned with the original video $\mathrm{V}$, the aligned \gls{tts} signal $\hat{x}$ turns out to be synchronized with the video itself.

We address this point using the \acrfull{dtw} implementation provided by Synctoolbox library~\cite{muller2021sync}. This toolbox integrates and combines several techniques for the given task, such as multiscale \gls{dtw}, memory-restricted \gls{dtw}, and high-resolution music synchronization.
The method used was initially proposed for synchronizing music, but we also tested its effectiveness in the case of speech. The \gls{dtw} process computes the chroma features of the analyzed tracks and warps them by bringing them into temporal correspondence.
The pipeline block inputs the original speech track $x$, together with the \gls{tts} track $s$, and outputs the warped signal $\hat{x}$.
In particular, $x$ is the target signal and $s$ is the one to be warped. In our pipeline, both $x$ and $s$ contain the exact text and the output $\hat{x}$ has the same length as the target $x$.

\begin{figure}
    \centering
    \includegraphics[width=.7\columnwidth]{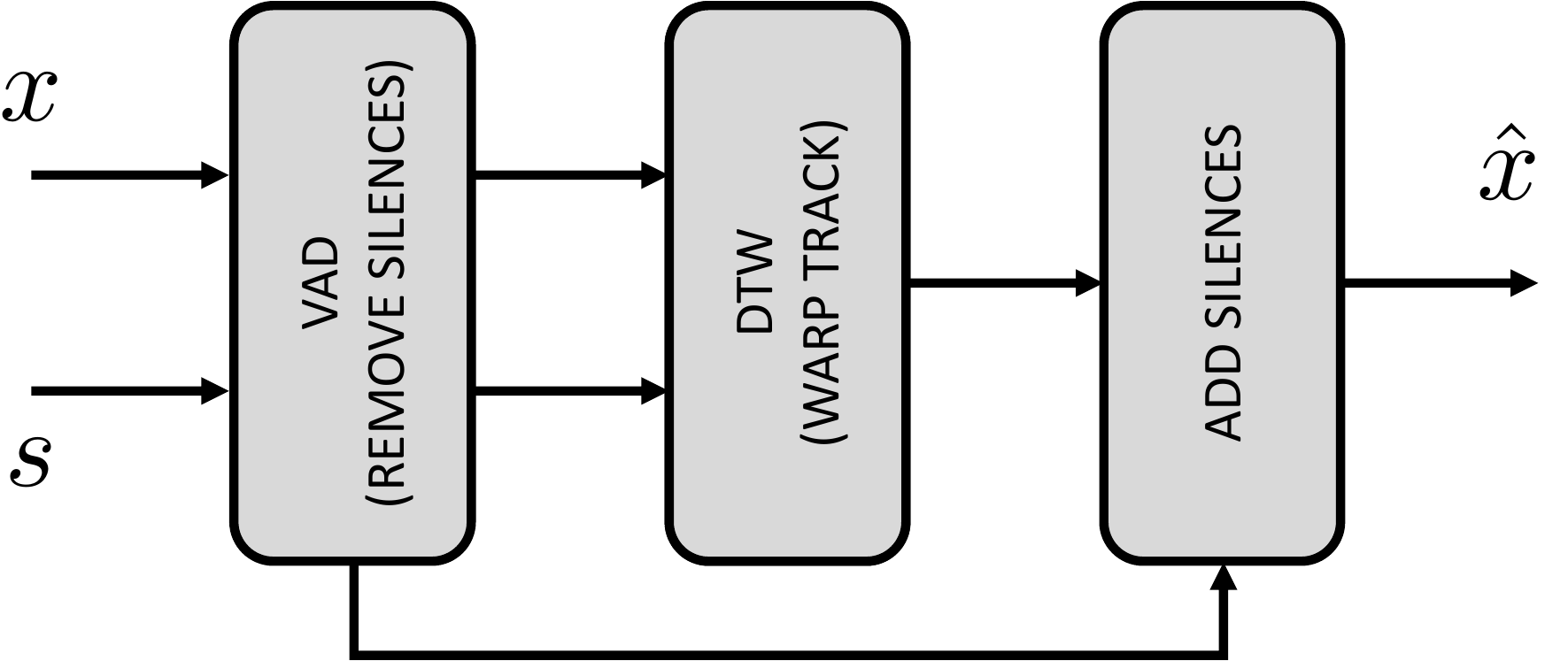}
    \caption{Pipeline of the speech-to-speech alignment block.}
    \label{fig:dtw}
\end{figure}

To improve the performance of this pipeline block we adopt a combined method of \gls{vad} + \gls{dtw}. In fact, in real cases, audio tracks often contain silences at their beginning or end, differently from \gls{tts} signals where silences are limited.
These silences can ruin the synchronization performances of the two tracks, as they are not symmetric.
To bypass this problem, we apply a \gls{vad} on both tracks before the alignment, removing the head and tail silences. Then, we perform the \gls{dtw} only on the voiced segments.
Finally, we add the silences removed from the target track to the warped one, obtaining a signal of the desired length.
This approach allows us to achieve more effective alignments and more realistic results.
Figure~\ref{fig:dtw} shows the complete pipeline of the alignment block.

\subsection{Post-processing}
\label{subsec:postprocessing}

Deepfake audio detectors generally perform very well when dealing with clean data, but their performance drops as these are post-processed.
When dealing with in-the-wild conditions, post-processing techniques are introduced to hide some artifacts present in the generated deepfake audio tracks.
For example, applying MP3 compression reduces the audio quality and hides some defects, while adding reverberation simulates the environment in which the audio was captured.
In our pipeline, we introduce a data augmentation block that allows us to generate more challenging data.
Table~\ref{tab:augmentation} shows the techniques we implemented and the parameters we considered for each transform, as shall be better explained in the next section.
We performed all the operations using the Python library audiomentations~\cite{audiomentations}.

%% file: 04_dataset_description.tex
\section{TIMIT-TTS Dataset Generation}
\label{sec:description}

This section provides all the details about the TIMIT-TTS dataset we release in this paper. After explaining its generation process, we illustrate its structure and possible applications.

\subsection{Reference dataset}

To generate a counterfeit speech dataset through the pipeline proposed in Section~\ref{sec:pipeline}, we need to define an audio-video set to use as a reference. 
Our goal is to produce a new version of the dataset where its audio component is replaced with a synthetic one.
Here we consider the VidTIMIT dataset~\cite{sanderson2002vidtimit, sanderson2009multi}.
This includes video and audio recordings of \num{43} people reciting \num{10} short sentences from the TIMIT Corpus~\cite{garofolo1993darpa}, for a total of \num{430} videos.
We chose this dataset for several reasons.
First, it is state-of-the-art and highly regarded within the scientific community.
Then, since the recorded sentences are extracted from the TIMIT Corpus, we are provided with all the transcripts of the video dialogues.
Therefore, we can avoid the text transcription step of the pipeline (see Figure~\ref{fig:pipeline}), which could introduce errors within the generated tracks, undermining the reliability of the released dataset.
This is crucial since the VidTIMIT recordings were done in an office using a broadcast quality digital video camera, resulting in noisy audio tracks that are difficult to transcript.
Moreover, the use of the official transcripts makes the generated speech perfectly synchronizable with the video, thus putting us in the most challenging forensic scenario where audio and video inconsistencies are minimal.
Finally, a counterfeited version of this dataset was released in \num{2018}.
This is called DeepfakeTIMIT~\cite{korshunov2018deepfakes} and includes \num{320} videos extracted from the VidTIMIT corpus modified using open-source software based on \glspl{gan} to create video deepfakes.
Being the released TIMIT-TTS an audio deepfake version of VidTIMIT, when it is used together with DeepfakeTIMIT, it provides audio-video content that is counterfeited in both modalities. This is extremely useful for the development of new multimodal deepfake detectors.

\subsection{Generated dataset}

To develop the TIMIT-TTS dataset, we consider the whole VidTIMIT corpus.
We generate a set of \num{430} synthetic speech tracks for each of the implemented generators, containing the same sentences as the reference videos. For the systems that support multispeaker synthesis, we synthesize a set of \num{430} tracks for each speaker.
We created several versions of the dataset corresponding to the different post-processing operations we apply to the generated speech tracks.
In particular, we consider two different processes: audio-video synchronization (\gls{dtw}) and data augmentation.
This results in the following four versions of the dataset:
\begin{itemize}[leftmargin=*]
    \item \textit{\textbf{clean\_data}}: all the synthetic audio tracks are clean and no post-processing is performed after the \gls{tts} generation process.
    \item \textit{\textbf{dtw\_data}}: \gls{dtw} is applied to the generated data. Each speech is synchronized with the corresponding video track from VidTIMIT.
    \item \textit{\textbf{aug\_data}}: data augmentation is applied to each speech track.
    \item \textit{\textbf{dtw\_aug\_data}}: both \gls{dtw} and data augmentation are applied to the generated data. First, we warp the tracks in time and then we augment them. We do so to prevent degradation from affecting the alignment process.
\end{itemize}

Considering the number of \gls{tts} methods and the number of speakers implemented, as shown in Table~\ref{tab:speakers}, each dataset partition is composed of \num{19 780} tracks, for a total of almost \num{80 000} speech signals on the entire dataset. All the tracks are released in \textit{wav} format considering a sampling rate of \SI{16}{\kilo\hertz}.
The complete dataset can be downloaded at this link\footnote{\url{https://zenodo.org/record/6560159}}.

Each partition of the dataset contains two splits, named \textit{single\_speaker} and \textit{multi\_speaker}. The first one includes all the tracks generated using \gls{tts} algorithms that support LJSpeech's speaker. The second includes the signals generated from the generators that implement speakers from datasets other than LJSpeech. Each of the two splits contains a subfolder for each generator, where the audio tracks are stored. The name of each track is \textit{dir\_track.wav}, where \textit{dir} and \textit{track} are respectively the names of the directories in which VidTIMIT is structured and of the tracks it contains. We adopted this naming to make it easy to link each deepfake audio track with its corresponding video.

Regarding data augmentation, we applied all the implemented techniques to each speech track, with a probability $p=0.3$ and a random value contained in a specific range for each method.
Following this application approach, some generated tracks will be edited with more than one method at a time, while others will remain clean.
At the same time, different augmentation levels will be considered for each track. This results in a dataset that is highly diverse and challenging to identify.
Table~\ref{tab:augmentation} shows all the augmentation techniques implemented, together with their considered ranges, while a list of the augmentation techniques applied to each signal can be found in a \textit{csv} file included in the partition folder.

\begin{table}[]
\centering
\caption{List of the implemented data augmentation techniques.}
\resizebox{.95\columnwidth}{!}{
\label{tab:augmentation}
\begin{tabular}{llc}
\toprule
\textbf{Augm. technique} &  \textbf{Parameter}        & \textbf{Application range}     \\ \midrule \midrule
Gaussian Noise         & $a$ - Amplitude    & $e^{-3}<a<1.5e^{-2}$ \\
Time Stretching        & $r$ - Rate         & $0.8<r<1.25$          \\
Pitch Shifting         & $s$ - Semitones    & $-8<s<8$              \\
High-pass Filtering    & $f$ - Cutoff freq. [Hz] & $20<f<2400$       \\
MP3 compression        & $a$ - Bitrate      & $8<b<64$              \\ \bottomrule
\end{tabular}}
\end{table}

The possible applications of TIMIT-TTS are numerous.
As regards synthetic speech detection, it is possible to perform that both in closed and open set scenarios. 
The high number of \gls{tts} generators implemented within the dataset allows us to include some of them in the train set while introducing others only in the test partition, making the classification task more challenging.
Furthermore, apart from binary classification, synthetic speech attribution can be performed.
This consists of a multi-class classification problem, where for each of the proposed tracks, it is required to find the \gls{tts} generation algorithm used to synthesize it.
Performing this study on TIMIT-TTS is fascinating since several of the proposed spectrogram generators only support the LJSpeech speaker. Indeed, synthetic speech attribution could be relatively easy to perform when each generator supports different speakers, but it becomes challenging when all the systems reproduce the same speaker. This type of analysis is presented in Section~\ref{sec:results}.

%% file: 05_results.tex
\section{Results and benchmarking}
\label{sec:results}

In this section, we benchmark the released dataset using objective metrics and show some of its possible applications, presenting the results obtained by testing it with state-of-the-art deepfake detectors.
We perform deepfake detection in both monomodal and multimodal scenarios, showing the effectiveness of considering multiple modalities at the same time.

\subsection{TIMIT-TTS statistics}

When generating synthesized audio data, many aspects need to be addressed to ensure the forged material is reliable and realistic.
These aspects include track length, silence duration, speech naturalness and number of supported speakers. Overlooking these aspects, we risk generating biased or easy-to-discriminate data.

The first aspect we analyze is the duration of the generated audio tracks.
As the dataset will be mainly used to develop deepfake detectors, we need the length of the audio tracks to be compatible with the window sizes used by most of the systems. Furthermore, we want to avoid differences between the duration of the signals generated with distinct \gls{tts} algorithms to prevent tracks generated by different methods from being easily discriminated.
The length of a signal generated through a \gls{tts} technique depends on the source text used as input. In our case, all the considered sentences are fixed and extracted from the TIMIT Corpus.

Table~\ref{tab:metrics} shows the duration values values for each \gls{tts} generation system. The average length over the entire dataset is equal to \SI{3.10}{\second}, while considering the single algorithms the durations range from \num{2.69} to \SI{3.82}{\second}.
The standard deviation between the duration of the different methods is not noticeable, being equal to \SI{0.33}{\second}. 
This means the length of the tracks does not constitute a discriminating element between the different generation algorithms, resulting in a reliable dataset.
When we apply \gls{dtw}, the average length of the tracks rises to \SI{4.25}{\second}. In this case, the average duration is the same for all generation algorithms. This is because the generated tracks have the same duration as the target ones extracted from VidTIMIT, so their length is fixed.

Secondly, we examined the length of silences contained in each track. 
Although silence is a fundamental component of speech, this is often overlooked in data generation, leading to biased tracks that are easy to discriminate~\cite{muller2021speech}.
This is a common problem, especially when dealing with \gls{tts} algorithms, where the prosodic component is less present~\cite{attorresi2022prosody} and the duration of the silences is shorter.
Table~\ref{tab:metrics} shows the silence durations of our tracks for both the original and the \gls{dtw} cases. 
Here we observe a higher difference between the algorithms, with duration values ranging between \num{0.05} and \SI{0.78}{\second}. However, when we apply \gls{dtw}, both the silence duration increase and the differences between the generation methods are reduced, homogenizing the synthesized data.

Next, as we are dealing with speech data, we assessed the naturalness of the generated tracks.
We do so to avoid releasing audio signals that sound too unrealistic.
We assume the \gls{mos} as a metric and compute it on the synthesized data through Mosnet~\cite{lo2019mosnet}.
\gls{mos} is a numerical measure of the human-judged overall quality of an event or experience, ranging from \num{1} (bad) to \num{5} (excellent). In our case, we use it to evaluate the naturalness of the generated speech tracks. 
The results for each generation algorithm are shown in Table~\ref{tab:metrics}.
We score an average \gls{mos} value greater than \num{3}, which is the threshold used to determine if a signal is acceptable or not. This means that, even if we are dealing with synthetic data, we are not neglecting the realism of the speech.
The application of \gls{dtw} has adverse effects on the \gls{mos} of the generated data, lowering the average computed on all the tracks by almost \num{0.2} points.

Finally, a crucial aspect to address in generated speech data is the number of supported speakers. As the primary goal of the TIMIT-TTS dataset is to perform binary detection of deepfakes, it is essential to provide several speakers.
Training a deepfake detector on a few speakers may make the model learn how to discriminate tracks based on the tone of voice they contain instead of the traces left by the \gls{tts} generators, as we will highlight in the following experiments.
Therefore, providing numerous speakers within the dataset helps avoid this bias and produce more effective models.

TIMIT-TTS implements a total of \num{69} different speakers in diverse numbers depending on the models used.
In addition to the LJSpeech voice, supported by numerous \gls{tts} generators, each multi-speaker system implements \num{8} different voices, \num{4} male and \num{4} female, from the VCTK or LibriSpeech datasets. The only exception is gTTS, which only supports \num{4} English voices. In this case, we have included all \num{4} within the dataset.
The number of speakers implemented for each generation method is shown in Table~\ref{tab:speakers}.

\begin{table}[]
\centering
\caption{Speech metrics for each \gls{tts} generator.}
\resizebox{.95\columnwidth}{!}{
\label{tab:metrics}
\begin{tabular}{lcccccc}
\toprule
\multirow{2}{*}{Generator} & \multicolumn{2}{c}{Track dur. [s]} & \multicolumn{2}{c}{Silences dur. [s]} & \multicolumn{2}{c}{MOS} \\ \cmidrule(lr){2-3}  \cmidrule(lr){4-5} \cmidrule(lr){6-7}
                           & Clean           & DTW          & Clean            & DTW            & Clean       & DTW       \\ \midrule \midrule
gTTS        & 3.82  & 4.25  & 0.55  & 1.29  & 3.59  & 3.39       \\
Tacotron    & 2.69  & 4.25  & 0.12  & 1.48  & 3.01  & 3.02     \\
GlowTTS     & 3.57  & 4.25  & 0.78  & 1.39  & 3.54  & 3.51      \\
FastPitch   & 2.74  & 4.25  & 0.41  & 1.47  & 3.48  & 3.35     \\
VITS        & 2.85  & 4.25  & 0.59  & 1.51  & 3.69  & 3.43      \\
FastSpeech2 & 3.03  & 4.25  & 0.05  & 1.32  & 3.03  & 3.00      \\
MixerTTS    & 3.35  & 4.25  & 0.07  & 1.32  & 3.04  & 3.02      \\
MixerTTS-X  & 3.34  & 4.25  & 0.11  & 1.35  & 3.02  & 3.01      \\
SpeedySpeech & 3.48 & 4.25  & 0.61  & 1.34  & 2.84  & 2.87      \\
Tacotron2   & 3.21  & 4.25  & 0.09  & 1.34  & 3.09  & 3.04      \\
TalkNet     & 3.02  & 4.25  & 0.05  & 1.33  & 3.00  & 2.99      \\
Silero      & 3.04  & 4.25  & 0.09  & 1.39  & 2.97  & 2.97      \\ \midrule 
Average     & 3.10  & 4.25  & 0.44  & 1.43  & 3.44  & 3.29      \\ \bottomrule 
\end{tabular}}
\end{table}

\subsection{Audio classification results}

To benchmark the generated data on the deepfake classification task, we consider an audio baseline that performs deepfake detection.
We adopt RawNet2~\cite{tak2021end}, a state-of-the-art end-to-end neural network that operates on raw waveforms.
It has been introduced to perform binary classification between real and fake data during the ASVspoof 2019 challenge~\cite{todisco2019asvspoof} and included as a baseline in the ASVspoof 2021 challenge~\cite{yamagishi2021asvspoof}.

Here we use the baseline for two different classification tasks. 
The first one is what it was initially proposed for, namely real vs. fake binary classification. The second is multiclass synthetic speech attribution. Here, given an audio signal generated with any \gls{tts} technique, we train the network to discriminate which algorithm has been used to synthesize the audio itself. 
For this second task, we modified the output layer of the network so that it contains many neurons equal to the number of classes we are addressing.
Although this is not the task for which the network was proposed, the problem is very close to that of deepfake classification and the considered model can address it without any issue~\cite{salvi2022exploring}. 
Also, the synthetic speech attribution problem has not yet been explored extensively, so there are not many networks proposed explicitly for the task. 
We illustrate all the experiments performed with RawNet2 in the following sections.

\subsubsection{Audio binary classification: synthetic speech detection}

In this experiment, we want to test how challenging the released dataset is in the deepfake detection task.
We perform binary classification considering the audio tracks of the VidTIMIT dataset as real and those of TIMIT-TTS as fake.
We use this dataset only in the test phase, following the approach presented in~\cite{muller2022does}, which is helpful for testing the generalization capabilities of a detector.
For this experiment we train RawNet2 on ASVspoof 2019, considering balanced classes and data augmentation on the training data.
We test the detector on the individual partitions of TIMIT-TTS.
When we consider augmented partitions, we also process real data from VidTIMIT following the same approach presented in the previous sections to make real and fake data as consistent as possible with each other.

\begin{figure}
    \centering
    \includegraphics[width=.8\columnwidth]{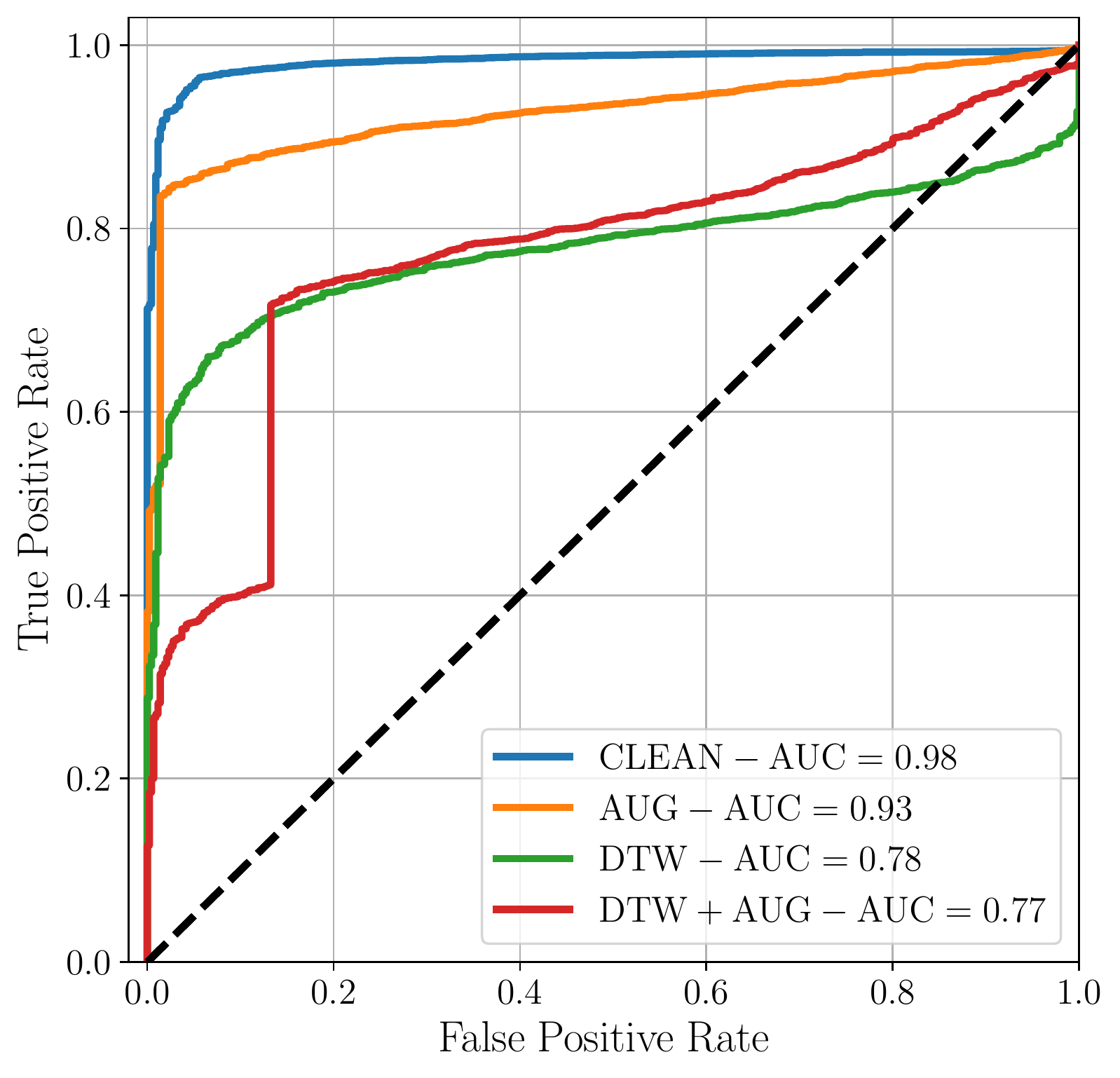}
    \caption{Audio binary classification - ROC Curves.}
    \label{fig:audio_ROC}
\end{figure}

\begin{figure}
    \centering
    \includegraphics[width=.8\columnwidth]{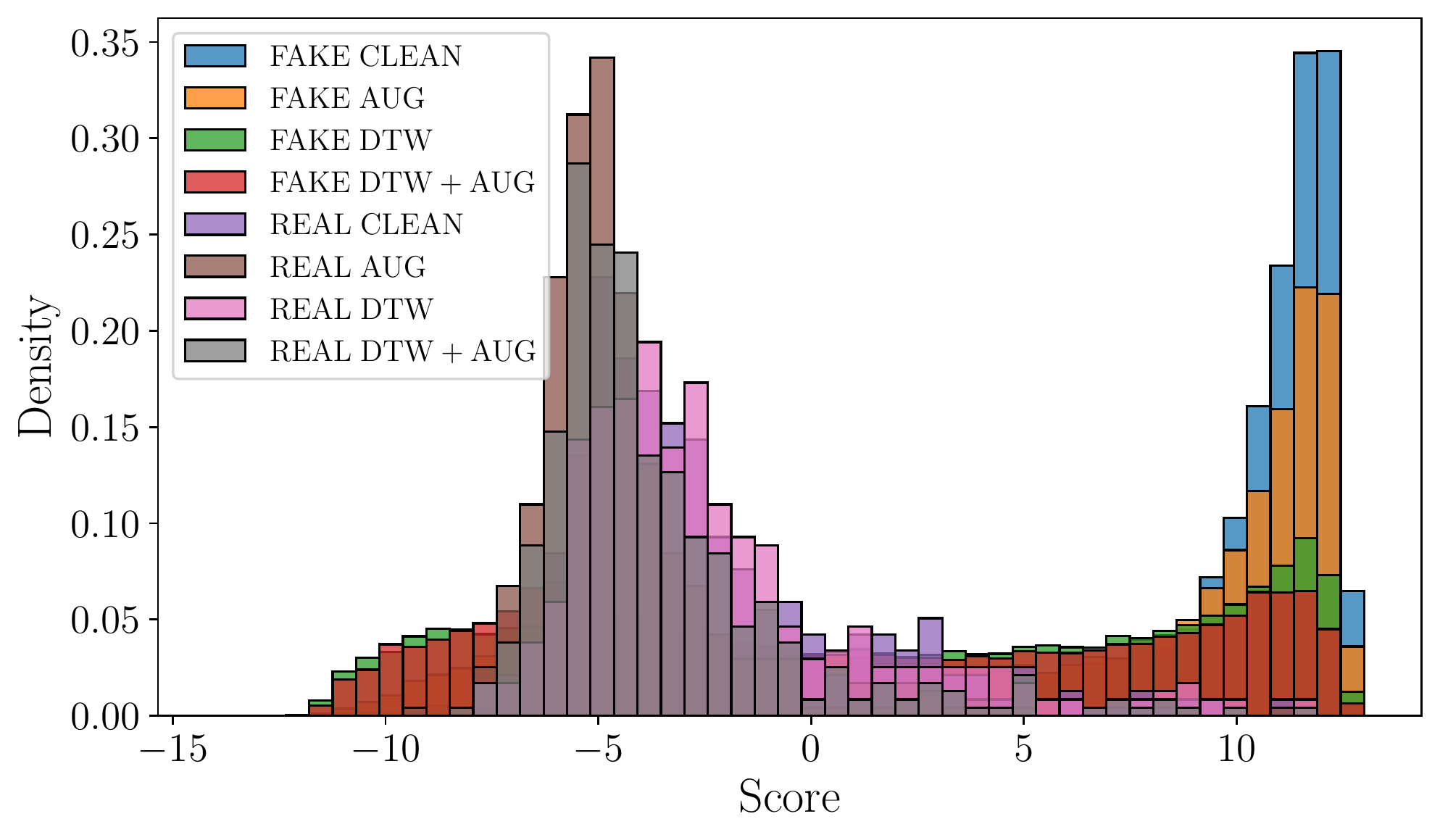}
    \caption{Audio binary classification - Scores distribution.}
    \label{fig:audio_hist}
\end{figure}

Figure~\ref{fig:audio_ROC} shows the results of the analysis by means of \gls{roc} curves and \gls{auc} values, while Figure~\ref{fig:audio_hist} shows the distributions of the scores for all the considered classes. In this case, higher scores mean higher confidence in identifying a track as fake.
We observe that the detection performance deteriorates as we increment the post-processing operations applied to the speech tracks.
In particular, the operation that degrades the accuracy the most is the speech-to-speech alignment, with an \gls{auc} value that drops by \num{0.20} between the clean and the \gls{dtw} cases.
This means that, although these tracks present a lower \gls{mos} value in Table~\ref{tab:metrics}, deepfake detectors must be explicitly trained on this type of data to discriminate them correctly.
Also, this shows that the detection problem of DTW tracks is not solved and our dataset could help in building new detectors that are more robust to in-the-wild conditions.
Finally, the augmented tracks are more challenging to detect than the clean ones, with an \gls{auc} value that drops by \num{0.05} between the two cases. As we mentioned above, such post-processing techniques hide some of the traces left by the \gls{tts} generators, making it more challenging to identify the artifacts present in the synthesized tracks.

\subsubsection{Audio multi-class classification: synthetic speech attribution}

In this experiment we want to test the TIMIT-TTS dataset on the synthetic speech attribution task.
This consists in identifying, given an input \gls{tts} track $y$, which algorithm has been used to synthesize it.
Formally, we have to determine $c_y \in \{c_1, c_2, ..., c_i\}$ where $i$ is the number of implemented \gls{tts} generation methods.
We consider all the \num{12} generation methods available in the TIMIT-TTS dataset, including all implemented speakers.
We split the corpus the into train and test sets following a \num{66}\% - \num{33}\% policy.
We ensure a coherent number of tracks for each generation algorithm in both the partitions.
We train the RawNet2 model for \num{100} epochs, using Cross Entropy as loss function and a learning rate equal to $10^{-4}$.

\begin{figure}
    \centering
    \includegraphics[width=.9\columnwidth]{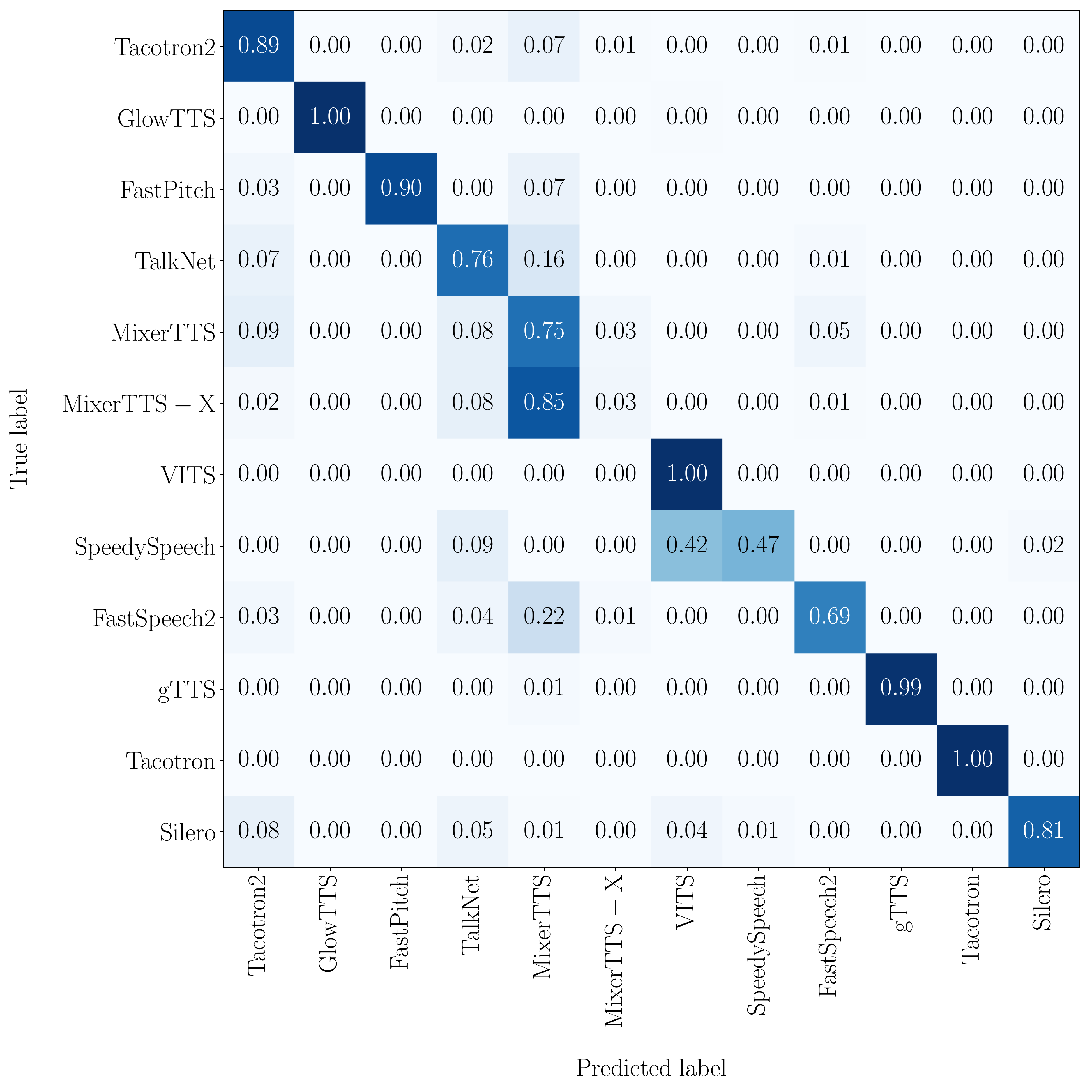}
    \caption{Confusion matrix showing the baseline performance on the synthetic speech attribution task, considering all the implemented \gls{tts} methods.}
    \label{fig:cm_multiclass}
\end{figure}

\begin{figure}
    \centering
    \includegraphics[width=.8\columnwidth]{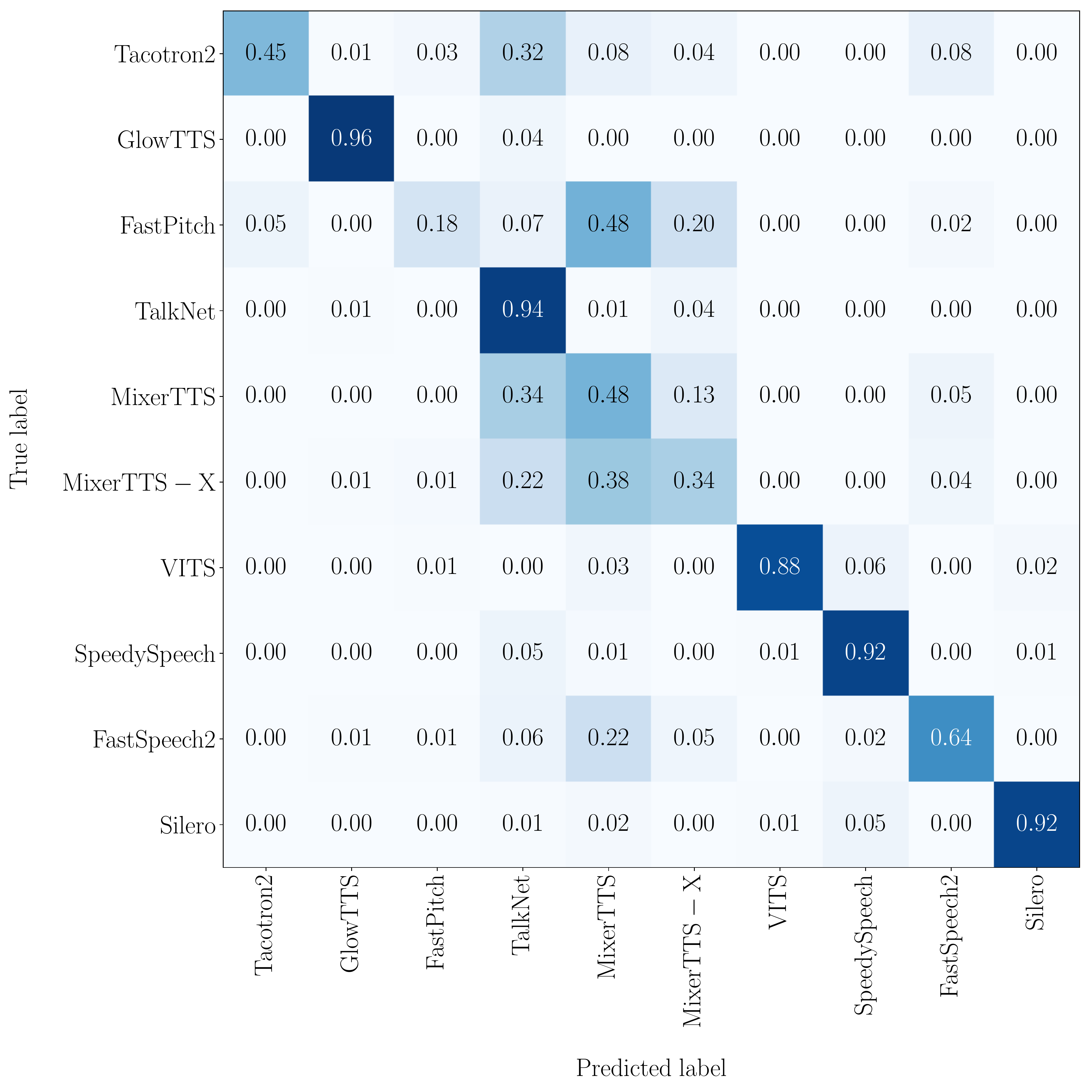}
    \caption{Confusion matrix showing the baseline performance on the synthetic speech attribution task, considering only the \gls{tts} methods trained on a single speaker.}
    \label{fig:cm_ljspeech}
\end{figure}

\begin{figure}
    \centering
    \includegraphics[width=.8\columnwidth]{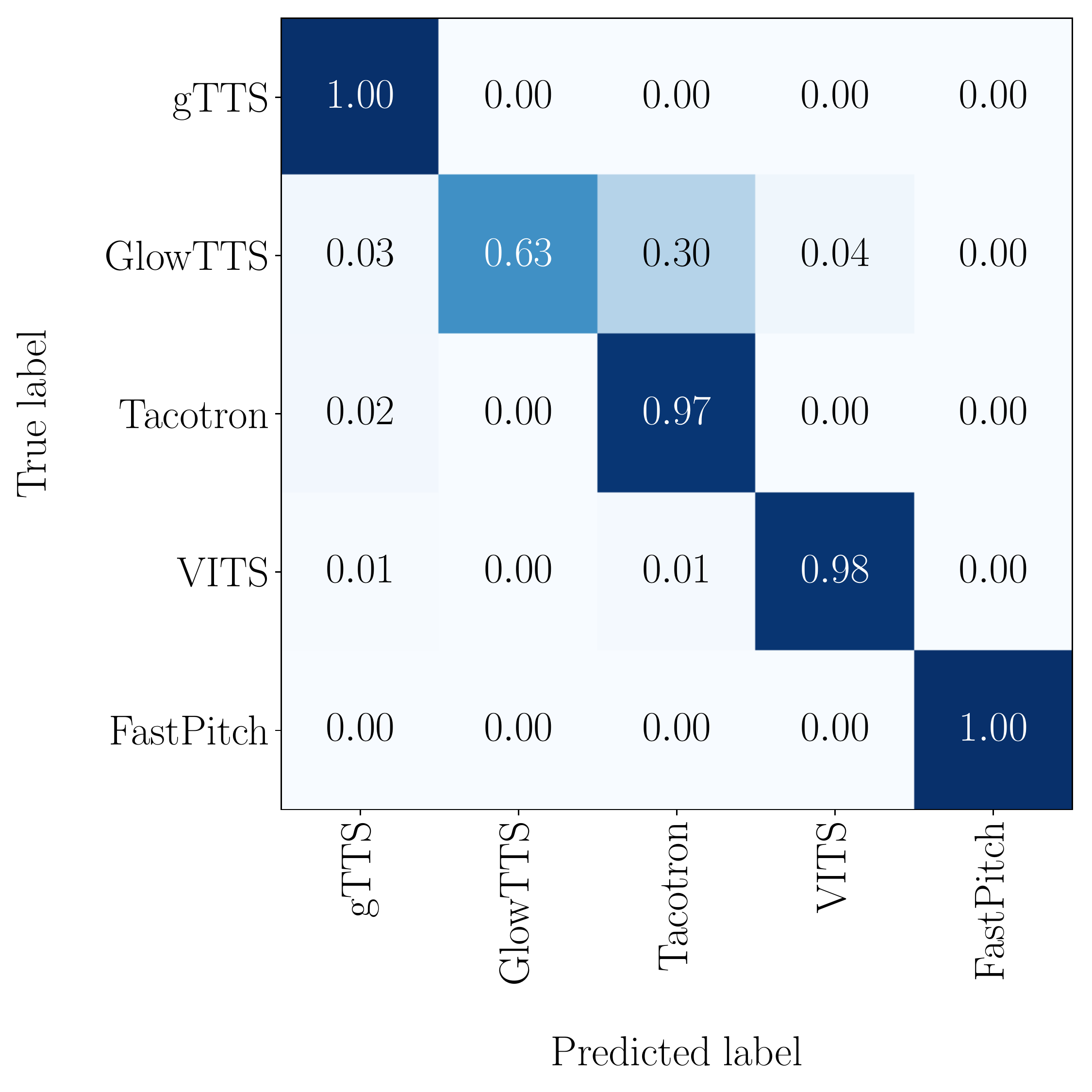}
    \caption{Confusion matrix showing the baseline performance on the synthetic speech attribution task, considering only the \gls{tts} methods that produce speech with multiple speakers.}
    \label{fig:cm_multi_speaker}
\end{figure}

Figure~\ref{fig:cm_multiclass} shows the results of the analysis through a confusion matrix.
We observe different performances for the considered algorithms. In particular, the systems trained to produce speech from multiple speakers are relatively easily identified, while those considering only one speaker are more challenging to distinguish.
This is due to the fact that the detection algorithm seems to leverage the different speakers to perform classification rather than focusing on the traces left by each \gls{tts} algorithm itself.
On the other hand, the methods that implement the same speaker force the model to learn how to discriminate tracks adequately, and the deterioration in performance is due to the difficulty of the required task.
To verify this hypothesis, we repeat the same experiment by independently considering the speech tracks generated by models trained on LJSpeech and those trained on other speakers. The results of this analysis are shown in Figures~\ref{fig:cm_ljspeech} and \ref{fig:cm_multi_speaker} and confirm the same trend as before, with the initial balanced accuracy value of \num{0.77} that drops from to \num{0.67} when we consider LJSpeech models and rises to \num{0.92} when considering the other models.
We believe this aspect is paramount when dealing with both deepfake detection and attribution tasks, as we do not want the results obtained by the algorithm to be biased by the considered speakers.
Indeed, having multiple \gls{tts} methods trained to reproduce the same voice constitutes a more challenging scenario as it forces the detector to learn the traces left by the generators. TIMIT-TTS, providing numerous generation methods trained on LJSpeech, can help develop new attribution algorithms.

\subsection{Video classification results}

As the final goal of our work is to use the proposed dataset to perform multimodal deepfake detection, we need to compare the final detection performance with those of the single modalities. For this reason, after analyzing the audio component, we operate the detection on the video one.
In this case we consider as baseline an EfficientNetB4~\cite{tan2019efficientnet} network modified by adding attention layers to improve its performance, following the implementation proposed in~\cite{bonettini2021video}.
As we did in the audio case, we consider a model trained on an external dataset to test its generalization capabilities.
We consider the model provided by the authors pre-trained on FaceForensics++~\cite{rossler2019faceforensics++} and test it on VidTIMIT and DeepfakeTIMIT datasets, considering them as real and fake data, respectively.

We build two different versions of the test set, corresponding to two different compression stages of the videos. In particular, we generate a high and low-quality version of the data obtained by considering two different values of quantization parameters (QP=\num{23} and QP=\num{40}), where higher QP means lower quality.
This has been done for two main reasons. First, this is the same compression approach considered in the FaceForensics++ dataset, so we used it to make our data comparable to those the model has been trained on.
Second, we want to study the robustness of the model to compression and analyze how much this influences the detection performance.
Robustness is a crucial aspect when dealing with deepfake detectors. The reason is that most of the multimedia material we deal with comes from social media, where they undergo several post-processing and compression steps. Developing a robust algorithm means being able to correctly analyze the multimedia material despite these operations.

\begin{figure}
    \centering
    \includegraphics[width=.8\columnwidth]{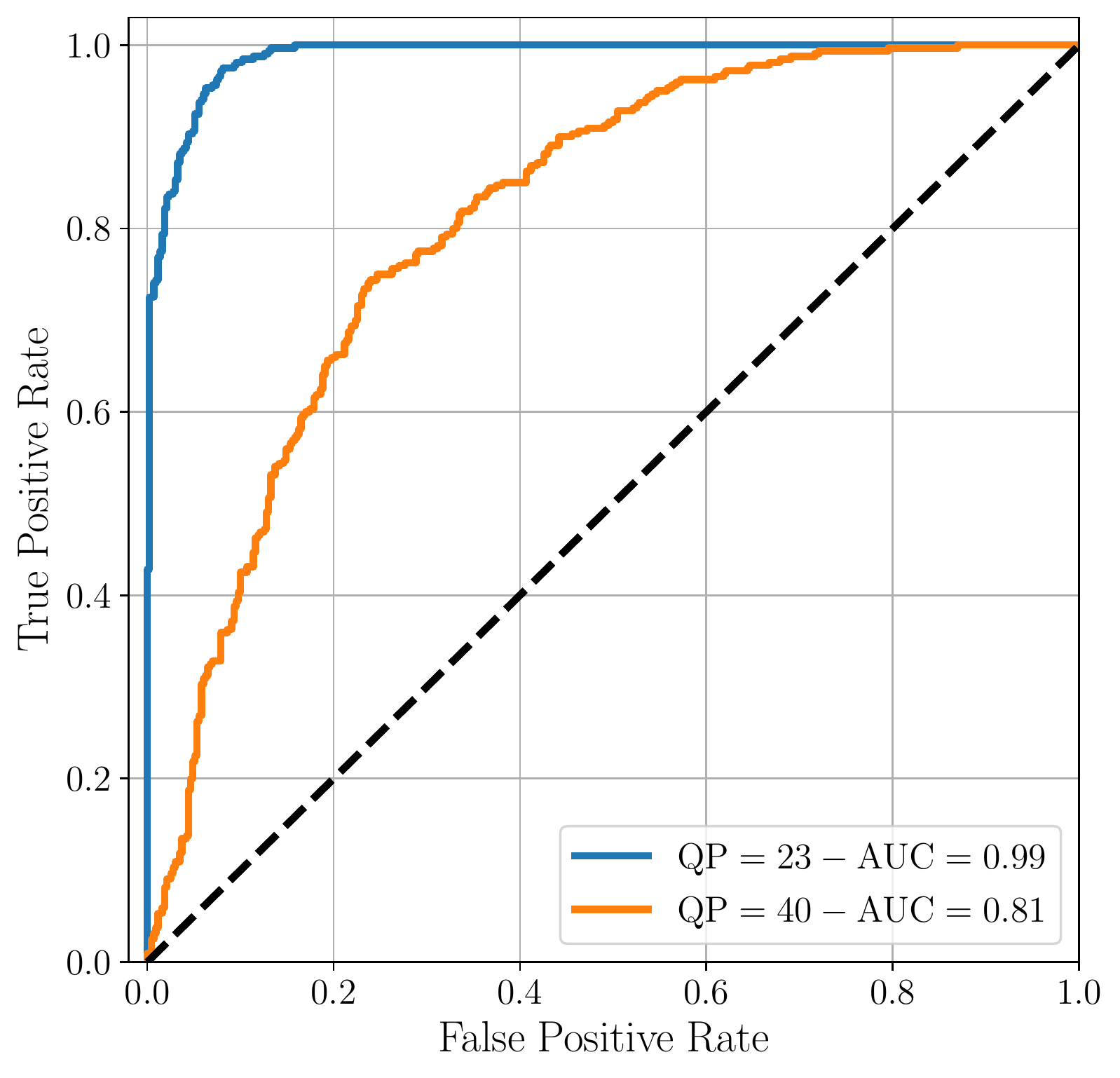}
    \caption{Video binary classification - ROC Curves.}
    \label{fig:video_ROC}
\end{figure}

\begin{figure}
    \centering
    \includegraphics[width=.8\columnwidth]{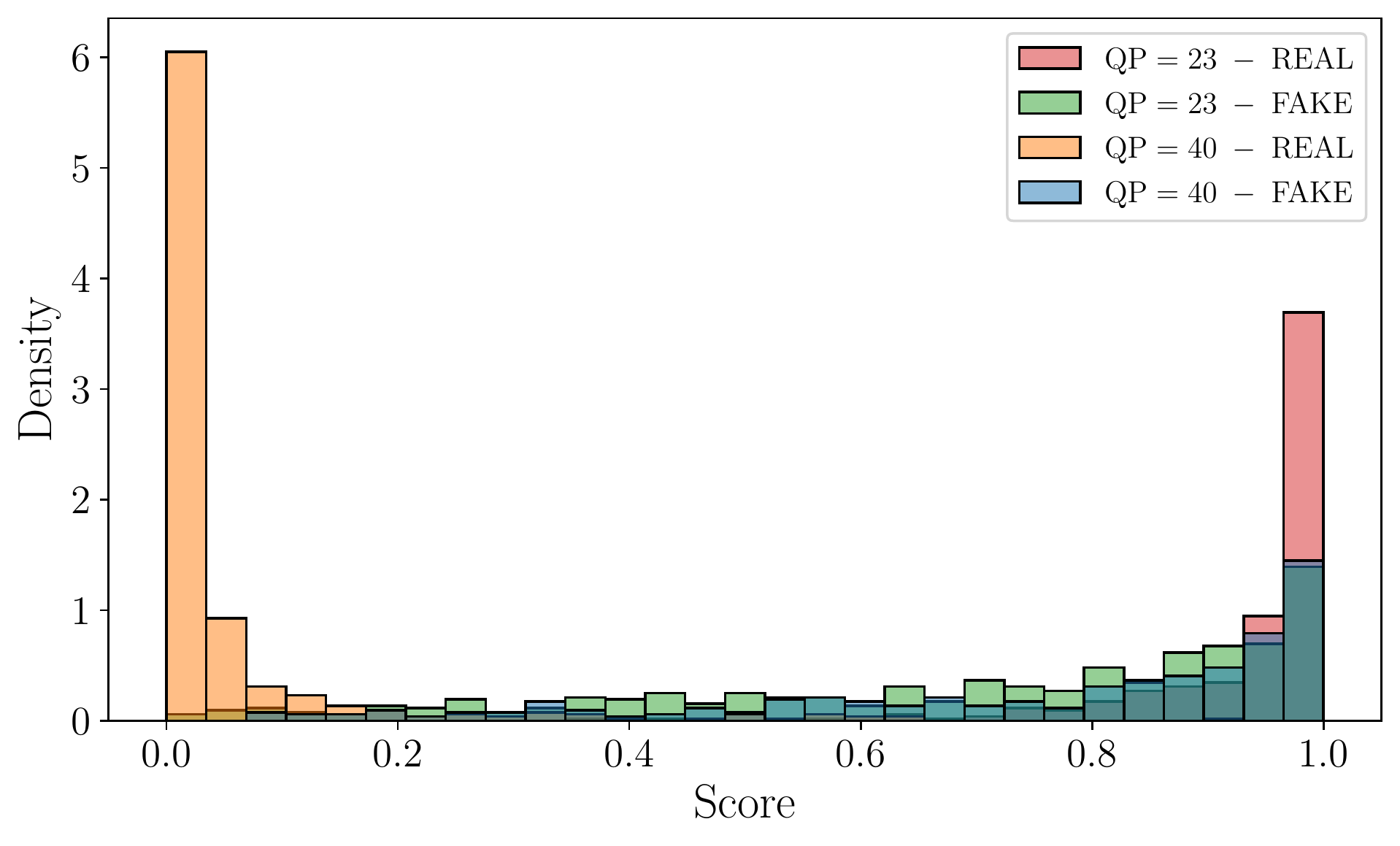}
    \caption{Video binary classification - Scores distribution.}
    \label{fig:video_hist}
\end{figure}

Figure~\ref{fig:video_ROC} shows the results of the detection task in terms of \gls{roc} curves and \gls{auc}, while Figure~\ref{fig:video_hist} shows the scores distributions in the considered cases. As in the previous experiment, higher score values mean a higher likelihood that the video is fake.
The detection task is accomplished very well when considering ``high quality'' videos, with an \gls{auc} value that is equal to \num{0.99}. 
This is a significant result, but we will unlikely find data with such high quality in in-the-wild conditions.
On the other hand, the performance significantly deteriorates when considering the ``low quality'' data, with an \gls{auc} value that drops by almost \num{0.2}. This leaves room for improvement in case of multimodal analysis.

\subsection{Multimodal classification results}

In this experiment we test the deepfake detection performance of the implemented baselines when considering a multimodal approach. We want to sample if simultaneously examining multiple aspects of a multimedia material can improve the detection capabilities or not.
To do so, we combine the VidTIMIT, DeepfakeTIMIT and TIMIT-TTS datasets and associate each audio track with its corresponding video.
In this way, we obtain a set of data that is falsified in both audio and video modalities.
During this study we analyze the two following scenarios:
\begin{itemize}
    \item \textbf{Scenario 1} - We only consider videos where both their modalities belong to the same class, e.g., audio and video are both real or both fake.
    \item \textbf{Scenario 2} - We consider videos where all the combinations between classes are possible, including data that are counterfeited in only one modality at a time. In this case, we label a video as fake when at least one between its audio and video components is falsified.
\end{itemize}
We do so to consider two different application cases for a multimodal approach. In the first scenario, since the classes of the two components are the same, it would also be possible to use a monomodal approach. Nonetheless, we show that analyzing different aspects of the given material can help improve the detection performance.

The second scenario, on the other hand, is more similar to a real-world cases. Here, using a multimodal approach is fundamental since analyzing only one aspect at a time we would lose information and have partial results. For example, we would be unable to detect videos that are counterfeited in just one modality if that is different from the one we are analyzing.
For both scenarios, we consider the baselines introduced above for the single modalities, and we fuse their score in two different ways.
In the first case, we compute the average between the two scores, while in the second we consider the higher of the two, which identifies the analyzed element as more likely to be false.

\begin{figure}
    \centering
    \includegraphics[width=.8\columnwidth]{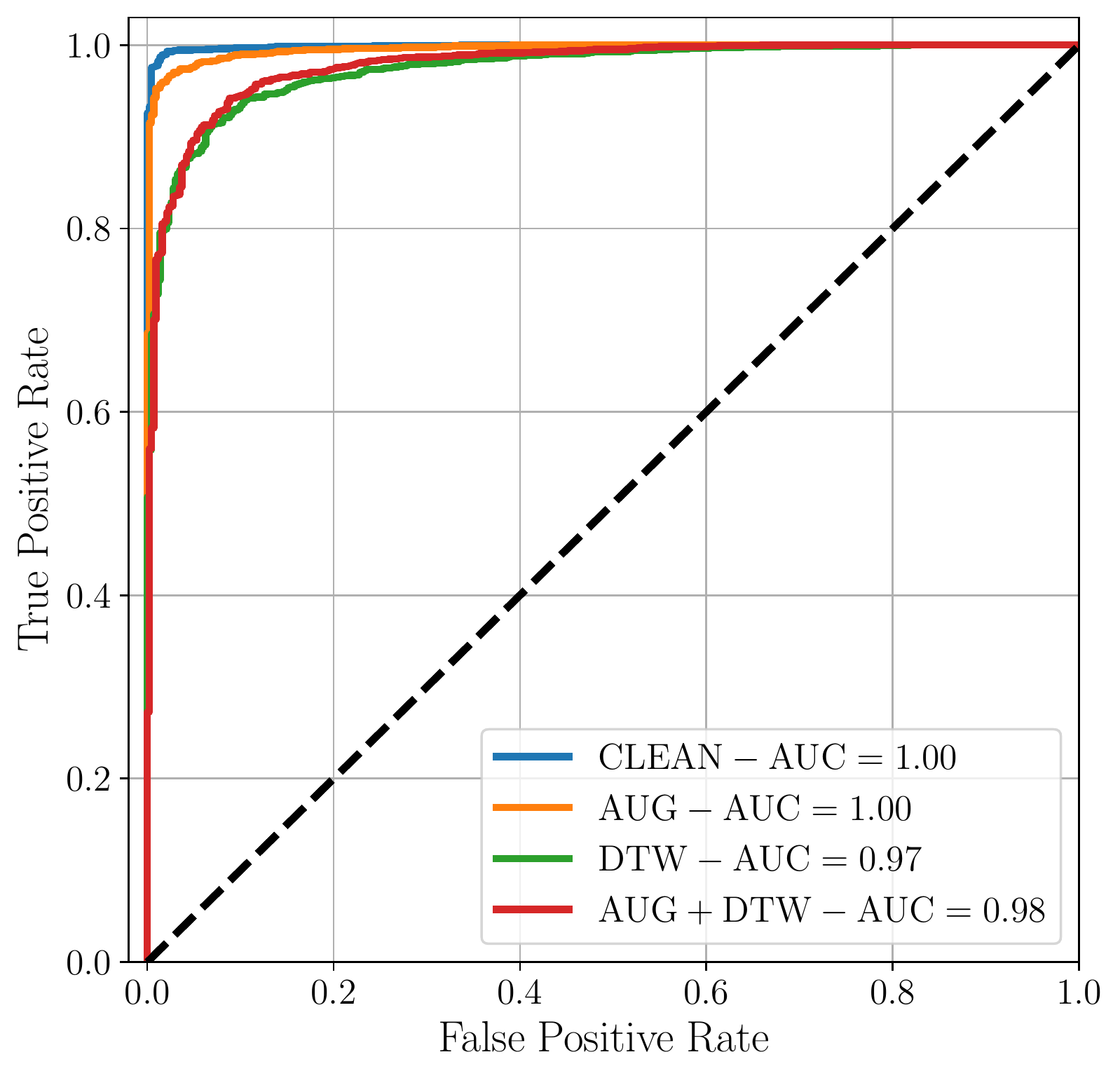}
    \caption{Multimodal binary classification - Scenario 1 (RR vs. FF) - QP=23.}
    \label{fig:multiclass_qp_23}
\end{figure}

\begin{figure}
    \centering
    \includegraphics[width=.8\columnwidth]{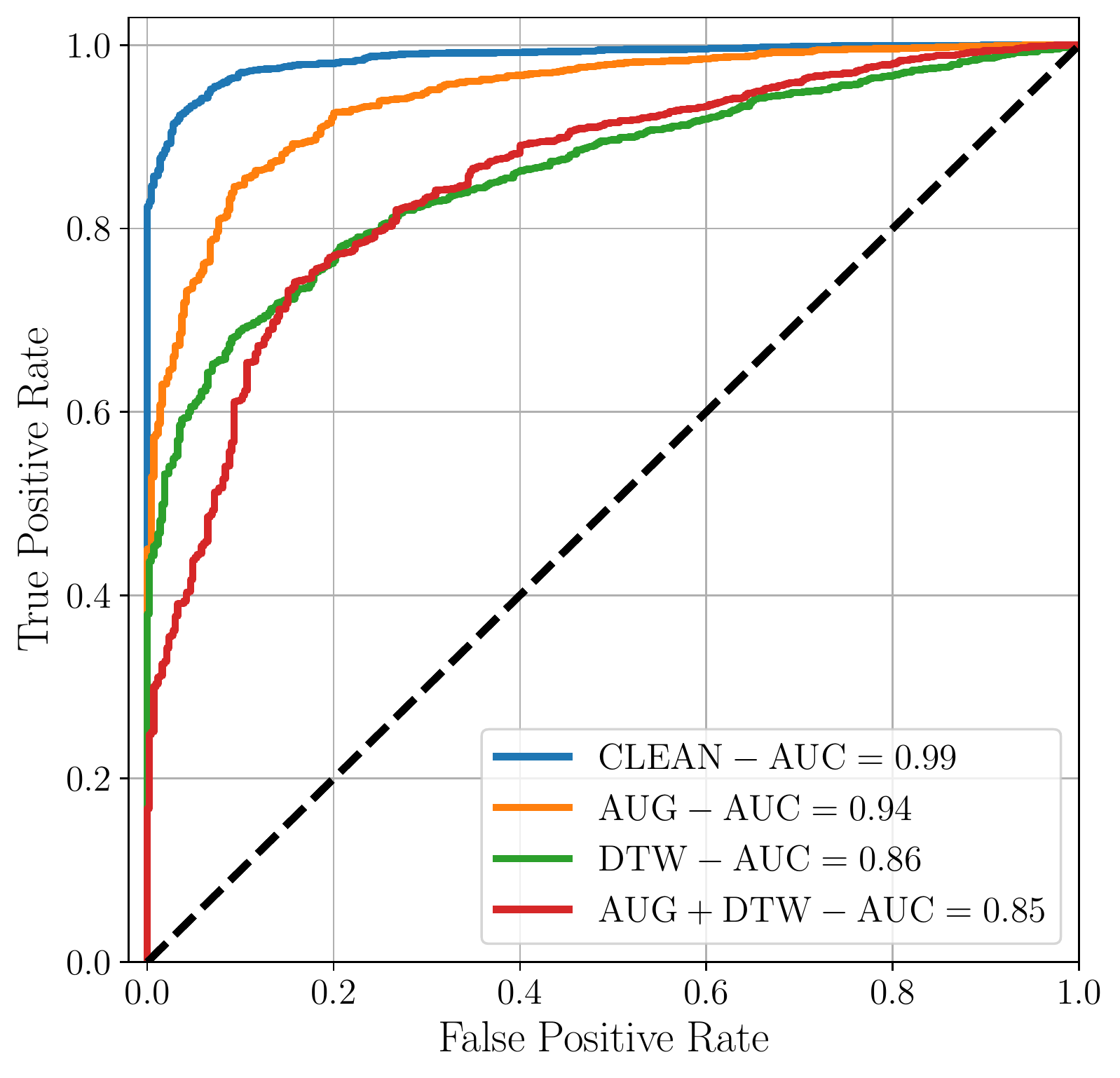}
    \caption{Multimodal binary classification - Scenario 1 (RR vs. FF) - QP=40.}
    \label{fig:multiclass_qp_40}
\end{figure}

The results conducted on the first scenario are shown in Figures~\ref{fig:multiclass_qp_23} and \ref{fig:multiclass_qp_40}, divided according to the compression applied to the video modality. The detection performance improves significantly, especially when dealing with post-processed data. In particular, the \gls{auc} values improve in all the cases compared to the corresponding monomodal experiments.
In the second scenario, likewise, the multimodal approach performs considerably better than the monomodal ones.

\begin{figure}
    \centering
    \includegraphics[width=.8\columnwidth]{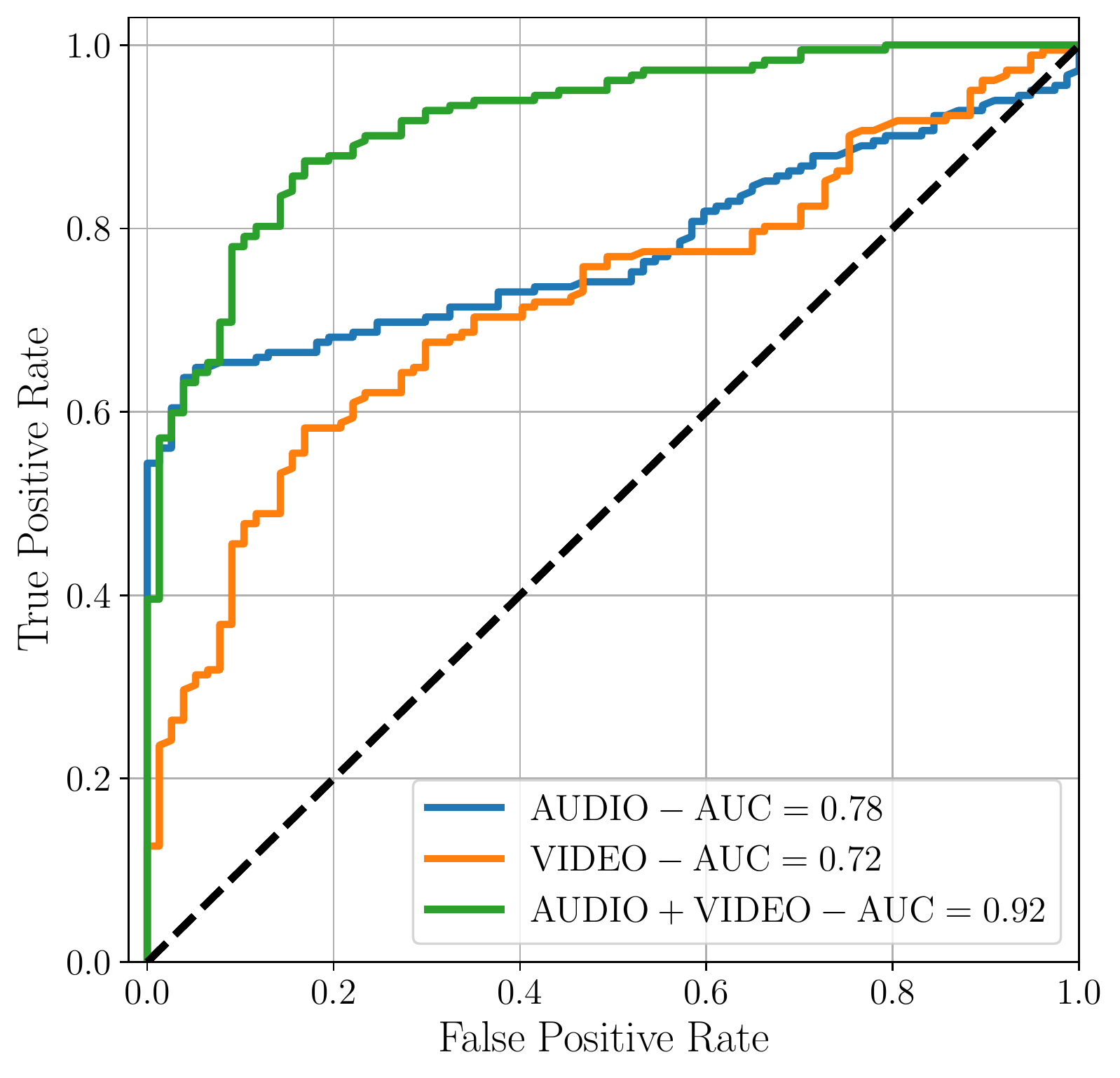}
    \caption{Multimodal binary classification - Scenario 2.}
    \label{fig:multiclass_all_classes}
\end{figure}

Figure~\ref{fig:multiclass_all_classes} shows the obtained results in the case we consider clean audio data and a QP=\num{23} for the video, where an \gls{auc} improvement of \num{0.15} is achieved over both the single modalities.
This is very interesting since it allows us to detect fake videos that we could not find otherwise.
We highlight that such positive results have been achieved by fusing the scores of the monomodal detectors in a very straightforward way. We are confident that combining them more smartly could further improve the performance, demonstrating the effectiveness of multimodal deepfake detectors.

%% file: 06_conclusion.tex
\section{Conclusion}
\label{sec:conclusion}

In this work we presented a pipeline to forge synthetic audio content starting from an input video in order to generate a multimodal deepfake dataset. We used this pipeline to generate and release TIMIT-TTS, a synthetic speech dataset that includes audio tracks generated using 12 different \gls{tts} systems, among the most advanced in the literature, for a total amount of almost \num{80 000} tracks.
The released dataset has several applications in the forensics field, such as synthetic speech detection and attribution.
Moreover, it can be used in conjunction with other well-established deepfake video datasets to perform multimodal studies, bridging an overlooked aspect in the current state-of-the-art.
From the presented results, it emerges that multimodal analyzes improve the performance of the detectors, producing more capable and robust systems. At the same time, however, the performances are not entirely satisfactory, so we need more multimodal deepfake datasets, like the one we release, to train and test the developed networks.

This is the dataset's first version, and future developments will be released. There are several aspects worth investigating and synthesis algorithms that have not been included in this set.
Regarding \gls{tts} systems, we want to examine the effects of using different vocoders on the performance of deepfake detectors and implement a higher number of speakers for all the systems.
Moreover, we also want to include \gls{vc} algorithms in the study since they have not been involved in this work.
Nonetheless, we hope this work will help the development of new multimodal deepfake detectors and provide new data to train and test existing systems to make them able to address in-the-wild conditions.